\documentclass[intlimits,twoside,a4paper]{article}

\usepackage[cp1251]{inputenc}

\usepackage[eqsecnum]{cmpj3}

\issue{2022}{25}{1}{13302}
\doinumber{10.5488/CMP.25.13302}

\title[Decoherence in open quantum systems]
{Decoherence in open quantum systems: influence of the intrinsic bath dynamics}
\author[V. V.~Ignatyuk, {V. G.~Morozov}]
{V. V.~Ignatyuk\orcid{0000-0003-4021-441X}\refaddr{label1}\thanks{Corresponding author: \email{ignat@icmp.lviv.ua}.},
       \framebox{V. G.~Morozov}\orcid{0000-0001-6507-6522}\refaddr{label2}\thanks{Prof.~{V.~G. Morozov} passed away soon after this article had been submitted to Condensed Matter Physics \cite{inMemoriam}. This is my last paper prepared in collaboration with my colleague and a good friend.}}

\addresses{
\addr{label1} Institute for Condensed Matter Physics, 1 Svientsitskii St., 79011 Lviv, Ukraine
\addr{label2} 
MIREA-Russian Technological University, 78 Vernadsky Ave., 119454 Moscow, Russia}

\Keywords{master equation, projection technique, dephasing model, decoherence}

\newcommand{\beq}{\begin{equation}}
\newcommand{\eeq}{\end{equation}}
\newcommand{\bi}{\begin{itemize}}
	\newcommand{\ei}{\end{itemize}}
\newcommand{\bea}{\begin{eqnarray}}
\newcommand{\eea}{\end{eqnarray}}
\newcommand{\ban}{\begin{eqnarray*}}
	\newcommand{\ean}{\end{eqnarray*}}
\newcommand{\barr}{\begin{array}}
	\newcommand{\earr}{\end{array}}

\date{Received October 13, 2021, in final form December 21, 2021}
\begin{document}

\maketitle

\begin{abstract}
The non-Markovian master equation for open quantum systems is obtained by gene\-ra\-li\-za\-tion of the standard Zwanzig-Nakajima (ZN) projection technique. To this end, a coupled chain of equations is written for the reduced density matrices of the bath $\varrho_{B}(t)$ and of the system $\varrho_{S}(t)$. A formal solution of the equation for $\varrho_{B}(t)$ in the 2nd approximation in interaction yields a specific extra term related to the intrinsic bath dynamics. This term is nonlinear in the reduced density matrix $\varrho_{S}(t)$, and vanishes in the Markovian limit. To verify the consistence and robustness of our approach, we apply the generalized ZN projection scheme to a simple dephasing model. We study the obtained kinetic equation both in the Markovian approximation and beyond it (for the term related to the intrinsic bath dynamics) and compare the results with the exact ones.

\printkeywords
%
\end{abstract}

\section{Introduction}\label{sec1}

When studying dynamical processes such as relaxation, decoherence, buildup of correlations due to the interaction of an open quantum system with its environment, at a certain stage of investigation one inevitably faces the question: do these phenomena exhibit the Markovian behaviour or not?
Many physical systems are believed to be described within Markovian approximation, since the coupling to the environment is weak (Born approximation), and/or the correlations in the bath decay fast on the characteristic time scale. However, there are situations, when the memory effects in the bath cannot be neglected, and the Markov assumption is not applicable anymore. This can be due to strong system-environment couplings \cite{4in-FB}, correlations and entanglement in the initial state \cite{6in-FB}, at the heat transport in nanostructures \cite{13in-FB}, or due to the specific character of the finite reservoirs \cite{9in-FB}.
The latter cases are of a particular interest, because the environment of an open quantum system due to its compactness frequently cannot be regarded as a thermal bath \cite{8in-FB}. In such a case, the dynamics of the reservoir $B$ should be treated (at least at the initial stage of evolution) on equal footing with the dynamics of the $S$-subsystem.

A powerful tool for dealing with such systems is provided
by the projection operator techniques \cite{14in-FB,15in-FB}, which was developed by Nakajima \cite{16in-FB}, Zwanzig \cite{17in-FB}, and Mori
\cite{18in-FB}. This approach has manifested its efficiency in the derivation of generalized master equations and in investigation of the non-Markovian dynamics in initially correlated open quantum systems \cite{Meier-Tannor-1999}, in the spin star systems \cite{Petruccione-2004} and many others.

However, the ZN  scheme has some disadvantages. Though the generic master equation is usually written down up to the 2nd order in the interaction, the time convolution in the kinetic kernels is determined by the \textit{full} evolution operators (including the interaction part $V$ of the total Hamiltonian). Thus, to provide a consistent analysis of the system dynamics, one should take into account
as many terms in the coupling constant as possible when expanding the master equation in the series in $V$. This is a cornerstone of the time-convolutionless equation \cite{BP-Book} and of the related projection technique, when one moves from the retarded dynamics to the equations local in time with the time-dependent generators and the subsequent systematic perturbative expansion scheme.

	In the decades that passed since the pioneering works of Nakajima and Zwanzig, the projection methods technique has been  significantly improved \cite{review}. The main efforts are usually directed toward the construction of correlated
	projection superoperators \cite{Ferraro2008,Semin2012}, which
	consider the relevant part of the dynamics as a correlated
	system-environment state, rather than a tensor product state
	$\varrho_{S}(t)\otimes\varrho_{B}$, which allows one to reproduce the exact solutions already
	in the lowest order in the interaction \cite{Fischer-Breuer-2007}. 
	In the very recent paper \cite{preprint2021}  
	the correlated projector technique has been generalized, yielding the so-called adapted projection operators (APO).
	The APO
	master equation encloses the full dependence on the initial
	correlations in a \textit{homogeneous} term, which is
	essential to avoid the problems \cite{BP-Book} dealing with an \textit{inhomogeneous} time 
	contribution arising from the unfactorizable initial states. An application of the APO methods manifested its efficiency in several cases, e.~g., in investigating  
	the polarization degrees of freedom of a photon
	going through a quartz plate --- the system, which can be described by a similar dephasing model. Another kind of projections, namely, the projections
	to the environment subspaces corresponding to a given
	energy, was proposed in \cite{Esposito2003}, which allowed the authors to derive an equation for the time-dependent reduced density matrix of the environment $\varrho_B(t)$ and to  closely approach the problem of taking  account of the intrinsic bath dynamics (BD).

However, none of the above methods considers the dynamics of the environment $B$ from the very beginning, on equal footing with that of an open quantum system $S$. Usually, the evolution equation for the reduced density matrix of the environment $\varrho_B(t)$ is neglected too. Though for the infinite thermal baths this simplification seems to be reasonable, its validity could be questioned, when one deals with the finite size reservoirs \cite{9in-FB} or tries to investigate the impact of the running correlations \cite{particles}. In other words, the open system dynamics should be definitely  in accord with the concept \cite{ZMR,Gemmer-Michel-Physica}, claiming that an evolution towards
local equilibrium is always accompanied by an increase of the system-bath correlations. 

In this paper, we generalize the standard ZN projection technique on our way to derive the non-Markovian master equation for open quantum systems. In section~\ref{sec2}, we start from the coupled set of equations for the reduced density matrices of the system $\varrho_S(t)$ and of the environment $\varrho_B(t)$. Then, in section~\ref{sec3}, we insert a formal solution of the dynamic equation for $\varrho_B(t)$ into the equation for the reduced density matrix of the $S$-subsystem and obtain a generalized master equation in the second order in  interaction. 
The above generalization gives rise to the extra term of a very specific structure:
i) it is nonlinear in $\varrho_S(t)$ and ii) vanishes in the Markovian limit. These two points make the situation very similar to what one faces when studying the inset of the running correlations: the generic kinetic equations are strongly nonlinear, and the correlational contribution to the collision integral tends to zero in the Markovian limit \cite{ZMR,MorRop01}. We also briefly compare our results with those following from the standard ZN scheme.

	To verify the consistency of our approach, in section~\ref{sec4}  we apply  the generalized projection scheme to a simple dephasing model \cite{PRA2012} and obtain the kinetic equation for the system coherence. It is pointed out that the application of the Markovian approximation (MA) allows us to reproduce the exact results reported in \cite{PRA2012} in the lowest orders in the interaction. On the other hand, going beyond the MA in the term dealt with the BD leads to the renormalization of the phase shift and the generalized decoherence. We demonstrate that in a certain domain of the state parameters, the above renormalization allows one to describe the correlational contribution to the generalized decoherence more precisely than the ZN scheme.

In the last section, we make conclusions and discuss the obtained results in their relation to the concept of the running correlations build-up.
\newpage

\section{Basic  equations}\label{sec2}

Suppose that the composed system ($S+B$) consists of the open quantum system (subsystem $S$) and its surrounding $B$, which usually can be considered as a thermal bath. The total Hamiltonian of this system
\beq
\label{Hamilt}
H(t)=H^{}_{S}(t) + H^{}_{B} + V
\eeq
consists of the term $H^{}_{S}(t)$ corresponding to the open quantum system $S$ (which in the general case is allowed to depend on time $t$ due to the action of the external fields), the summand $H_{B}$ related to the bath, and the interaction term $V$.

The density matrix $\varrho(t)$ of the composed system obeys the quantum Liouville equation (hereafter we put $\hbar=1$),
\beq
\label{Eq-rho}
\frac{\partial\varrho(t)}{\partial t}= -\ri\left[H(t),\varrho(t)\right].
\eeq

For further convenience, let us pass to the interaction picture $\widetilde{\varrho}(t)$ for the total density matrix, which is defined as
\beq
\label{rho-IP}
\widetilde{\varrho}(t)= U^{\dagger}(t)\varrho(t)U(t),
\eeq
where we have introduced the unitary evolution operators
\beq
\label{U-def}
\begin{array}{c}
	U(t)=U^{}_{S}(t)U^{}_{B}(t),
	\\[10pt]
	\displaystyle
	U^{}_{S}(t)= \exp^{}_{+}\left\{-\ri\int^{t}_{0} \rd t'\,H^{}_{S}(t')\right\},
	\qquad
	U^{}_{B}(t)=\exp\left(-\ri tH^{}_{B}\right).
\end{array}
\eeq
In equation~(\ref{U-def}), the expression $\exp_+\left\{\ldots\right\}$ denotes the time ordered exponent, which takes a much simpler form $U^{}_{S}(t)=\exp\left(-\ri tH^{}_{S}\right)$ if the open system Hamiltonian $H_S$ does not depend on time. The unitary operator $U(t)$ and its Hermitian conjugate counterpart $U^{\dagger}(t)$ obey the evolution equations
\beq
\label{U-eq}
\frac{\rd U(t)}{\rd t}=
-\ri\left\{H^{}_{S}(t)+H^{}_{B}\right\}U(t),
\qquad
\frac{\rd U^{\dagger}(t)}{\rd t}=\ri\, U^{\dagger}(t)\left\{H^{}_{S}(t)+H^{}_{B}\right\},
\eeq
wherefrom it is easy to present the quantum Liouville equation (\ref{Eq-rho}) in the interaction picture,
\beq
\label{Eq-rho-IP}
\frac{\partial\widetilde{\varrho}(t)}{\partial t}=
-\ri\left[\widetilde{V}(t),\widetilde{\varrho}(t)\right].
\eeq
Hereafter, operators in the interaction picture are defined as
\beq
\label{A-IP}
\widetilde{A}(t)= U^{\dagger}(t)A U(t).
\eeq
If $A^{}_{S}$ ($A^{}_{B}$) is an operator acting in the Hilbert space
of the subsystem $S$ (bath $B$), then the corresponding interaction representation can be written down as follows:
\beq
\label{A-IP-SB}
\widetilde{A}^{}_{S}(t)=U^{\dagger}_{S}(t)A^{}_{S} U^{}_{S}(t),
\qquad
\widetilde{A}^{}_{B}(t)=U^{\dagger}_{B}(t)A^{}_{B} U^{}_{B}(t).
\eeq
The reduced density matrices for the system $S$ and the
environment are introduced in a usual way,
\beq
\label{rho-SB}
\varrho^{}_{S}(t)= \text{Tr}^{}_{B}\varrho(t),
\quad
\varrho^{}_{B}(t)= \text{Tr}^{}_{S}\varrho(t),
\eeq
by taking a trace over the environment  (the system) variables.
The corresponding time averages for operators $A_S$ ($A_B$) can be introduced as follows:
\beq
\label{Ave-rho-SB}
\begin{array}{l}
	\langle A^{}_{S}\rangle^{t}\equiv
	\text{Tr}^{}_{SB}\left\{A^{}_{S}\varrho(t)\right\}=
	\text{Tr}^{}_{S}\left\{A^{}_{S}\varrho^{}_{S}(t)\right\},\quad
	\langle A^{}_{B}\rangle^{t}\equiv
	\text{Tr}^{}_{SB}\left\{A^{}_{B}\varrho(t)\right\}=
	\text{Tr}^{}_{B}\left\{A^{}_{B}\varrho^{}_{B}(t)\right\}.
\end{array}
\eeq

The reduced density matrix $\widetilde{\varrho}^{}_{S}(t)$ of the open quantum system is defined as
\beq
\label{IP-rho-S}
\widetilde{\varrho}^{}_{S}(t)=
U^{\dagger}_{S}(t)\varrho^{}_{S}(t)U^{}_{S}(t).
\eeq
Analogously, in the interaction picture, the reduced density matrix $\widetilde{\varrho}^{}_{B}(t)$ of the environment reads
\beq
\label{IP-rho-B}
\widetilde{\varrho}^{}_{B}(t)=
U^{\dagger}_{B}(t)\varrho^{}_{B}(t)U^{}_{B}(t).
\eeq
It is obvious that
$\widetilde{\varrho}^{}_{S}(t)=
\text{Tr}^{}_{B}\widetilde{\varrho}(t)$ and
$ \widetilde{\varrho}^{}_{B}(t)=
\text{Tr}^{}_{S}\widetilde{\varrho}(t)$.

To derive the evolution equations for the operators $\widetilde{\varrho}_S(t)$ and $\widetilde{\varrho}_B(t)$, we apply the following decomposition for the total density matrix:
\beq
\varrho(t)= \varrho^{}_{S}(t)\varrho^{}_{B}(t)+ \Delta \varrho(t),
\label{rho-decomp}
\eeq
where the correlation term satisfies the relations
\beq
\text{Tr}^{}_{S}\,\Delta \varrho(t)=0,
\quad
\text{Tr}^{}_{B}\,\Delta \varrho(t)=0.
\label{Del-rho-prop}
\eeq
In the interaction picture, equations~(\ref{rho-decomp})--(\ref{Del-rho-prop}) are converted into the similar relations,
\beq
\widetilde{\varrho}(t)=
\widetilde{\varrho}^{}_{S}(t) \widetilde{\varrho}^{}_{B}(t)
+ \Delta  \widetilde{\varrho}(t)
\label{rho-decomp1}
\eeq
and
\beq
\text{Tr}^{}_{B}\,\Delta  \widetilde{\varrho}(t)=0,
\qquad
\text{Tr}^{}_{S}\,\Delta  \widetilde{\varrho}(t)=0.
\label{IP-Del-rho-prop}
\eeq

Taking the trace $\text{Tr}^{}_{B}$ of both sides of
equation~(\ref{Eq-rho-IP}), we get
\beq
\label{IP-eq-S-1}
\frac{\partial\widetilde{\varrho}^{}_{S}(t)}{\partial t}=
-\ri\,\text{Tr}^{}_{B}\left[\widetilde{V}(t),\widetilde{\varrho}(t)\right].
\eeq
Using the decomposition (\ref{rho-decomp1}), one can rewrite equation~(\ref{IP-eq-S-1}) as
\beq
\label{IP-eq-S-2}
\frac{\partial\widetilde{\varrho}^{}_{S}(t)}{\partial t}=
-\ri\left[\widetilde{V}^{}_{S}(t),\widetilde{\varrho}^{}_{S}(t)\right]
-\ri\,\text{Tr}^{}_{B}
\left[\widetilde{V}(t),\Delta\widetilde{\varrho}(t)\right],
\eeq
where
\beq
\label{H-int-S}
\widetilde{V}^{}_{S}(t)=\text{Tr}^{}_{B}
\left\{
\widetilde{V}(t)\widetilde{\varrho}^{}_{B}(t)
\right\}.
\eeq

In a similar way, taking the trace $\text{Tr}^{}_{S}$ of equation~(\ref{Eq-rho-IP}), one obtains
\beq
\label{IP-eq-B-2}
\frac{\partial\widetilde{\varrho}^{}_{B}(t)}{\partial t}=
-\ri\left[\widetilde{V}^{}_{B}(t),\widetilde{\varrho}^{}_{B}(t)\right]
-\ri\,\text{Tr}^{}_{S}
\left[\widetilde{V}(t),\Delta\widetilde{\varrho}(t)\right],
\eeq
where
\beq
\label{H-int-B}
\widetilde{V}^{}_{B}(t)=\text{Tr}^{}_{S}
\left\{
\widetilde{V}(t)\widetilde{\varrho}^{}_{S}(t)
\right\}.
\eeq

To derive the equation of motion for the correlational part $\Delta\widetilde{\varrho}(t)$ of the total density matrix, 
let us rewrite equations~(\ref{Eq-rho-IP}), (\ref{IP-eq-S-2}), and
(\ref{IP-eq-B-2}) in the form
\beq
\label{IP-eq-rho}
\frac{\partial \widetilde{\varrho}(t)}{\partial t}=
-\ri{\mathcal L}(t)\widetilde{\varrho}(t),
\eeq
\beq
\label{IP-eq-S-L}
\frac{\partial \widetilde{\varrho}^{}_{S}(t)}{\partial t}=
-\ri{\mathcal L}^{}_{S}(t)\widetilde{\varrho}^{}_{S}(t)
- \text{Tr}^{}_{B}\left\{\ri{\mathcal L}(t)\,
\Delta \widetilde{\varrho}(t)\right\},
\eeq
\beq
\label{IP-eq-B-L}
\frac{\partial \widetilde{\varrho}^{}_{B}(t)}{\partial t}=
-\ri{\mathcal L}^{}_{B}(t)\widetilde{\varrho}^{}_{B}(t)
- \text{Tr}^{}_{S}\left\{\ri{\mathcal L}(t)\,
\Delta \widetilde{\varrho}(t)\right\}.
\eeq
Here, we have introduced the Liouville operators ${\mathcal L}(t)$, ${\mathcal L}_S(t)$, and ${\mathcal L}_B(t)$ via the corresponding commutators:
\beq
\label{L-def}
{\mathcal L}(t)A=[\widetilde{V}(t),A],
\qquad
{\mathcal L}^{}_{S}(t)A=[\widetilde{V}^{}_{S}(t),A],
\qquad
{\mathcal L}^{}_{B}(t)A=[\widetilde{V}^{}_{B}(t),A].
\eeq
Using the decomposition (\ref{rho-decomp1}) and taking into account equations~(\ref{IP-eq-rho})--(\ref{IP-eq-B-L}), one can obtain the equation of motion for the correlation contribution to the total density matrix,
\bea
\label{eq-del-rho}
\frac{\partial \,\Delta\widetilde{\varrho}}{\partial t}=
-\ri{\mathcal L}\widetilde{\varrho}
-\widetilde{\varrho}^{}_{S}
\left\{
-\ri{\mathcal L}^{}_{B}\widetilde{\varrho}^{}_{B}
- \text{Tr}^{}_{S}\left\{\ri{\mathcal L}\,
\Delta \widetilde{\varrho}\right\}
\right\}
-\widetilde{\varrho}^{}_{B}
\left\{
-\ri{\mathcal L}^{}_{S}\widetilde{\varrho}^{}_{S}
- \text{Tr}^{}_{B}\left\{\ri{\mathcal L}\,
\Delta \widetilde{\varrho}\right\}
\right\}.
\eea
The above equation can be rewritten in a more compact form,
\beq
\label{IP-eq-Del-rho}
\left(
\frac{\partial}{\partial t}
+{\mathcal Q}(t)\ri{\mathcal L}(t){\mathcal Q}(t)
\right) \Delta \widetilde{\varrho}(t)=
-{\mathcal Q}(t)\ri{\mathcal L}(t)
\widetilde{\varrho}^{}_{S}(t)\widetilde{\varrho}^{}_{B}(t),
\eeq
using the superoperators
\beq
\label{P-Q}
{\mathcal Q}(t)=1-{\mathcal P}(t),
\qquad
{\mathcal P}(t)A=
\widetilde{\varrho}^{}_{S}(t)\,\text{Tr}^{}_{S}A
+ \widetilde{\varrho}^{}_{B}(t)\,\text{Tr}^{}_{B}A.
\eeq
It can be shown that, when acting on operators with
zero trace, $\text{Tr}A=0$, the superoperator ${\mathcal P}(t)$
satisfies the relation ${\mathcal P}^{2}(t)={\mathcal P}(t)$, i.e., it is a
projector. 
A formal solution of equation~(\ref{IP-eq-Del-rho}) is
\beq
\label{Del-rho-formal}
\Delta\widetilde{\varrho}(t)=
- \ri\int^{t}_{0}\rd t'\,
{\mathcal U}(t,t'){\mathcal Q}(t'){\mathcal L}(t')
\widetilde{\varrho}^{}_{S}(t')\widetilde{\varrho}^{}_{B}(t'),
\eeq
where the superoperator ${\mathcal U}(t,t')$ satisfies the equation of motion
\beq
\label{MU-eq}
\frac{\partial {\mathcal U}(t,t')}{\partial t}=
-\ri{\mathcal Q}(t){\mathcal L}(t){\mathcal Q}(t){\mathcal U}(t,t'),
\qquad {\mathcal U}(t',t')=1,
\eeq
and has an explicit form
\beq
\label{MU-expr}
{\mathcal U}(t,t')=\exp^{}_{+}
\left\{
-\ri\int^{t}_{t'}\rd\tau\,
{\mathcal Q}(\tau){\mathcal L}(\tau){\mathcal Q}(\tau)
\right\}.
\eeq
Substituting expression (\ref{Del-rho-formal}) into
equations~(\ref{IP-eq-S-L}) and (\ref{IP-eq-B-L}),
we arrive at the closed system of equations
for the reduced density matrices,
\beq
\label{IP-eq-S}
\frac{\partial \widetilde{\varrho}^{}_{S}(t)}{\partial t}=
-\ri{\mathcal L}^{}_{S}(t)\widetilde{\varrho}^{}_{S}(t)
-\int^{t}_{0}\rd t'\,
\text{Tr}^{}_{B}
\left\{
{\mathcal L}(t){\mathcal U}(t,t'){\mathcal Q}(t'){\mathcal L}(t')
\widetilde{\varrho}^{}_{S}(t')\widetilde{\varrho}^{}_{B}(t')
\right\},
\eeq
\beq
\label{IP-eq-B}
\frac{\partial \widetilde{\varrho}^{}_{B}(t)}{\partial t}=
-\ri{\mathcal L}^{}_{B}(t)\widetilde{\varrho}^{}_{B}(t)
-\int^{t}_{0}\rd t'\,
\text{Tr}^{}_{S}
\left\{
{\mathcal L}(t){\mathcal U}(t,t'){\mathcal Q}(t'){\mathcal L}(t')
\widetilde{\varrho}^{}_{S}(t')\widetilde{\varrho}^{}_{B}(t')
\right\}.
\eeq
However, equation~(\ref{IP-eq-S}) cannot be considered yet as a master equation for the open quantum system since it depends on the reduced density matrix of the environment. Solving equation~(\ref{IP-eq-B}) with respect to the density matrix of the environment $\widetilde{\varrho}_B(t)$ seems unrealistic. 
Thus, the only way is to use some approximations for $\widetilde{\varrho}_B(t)$ obtainable from equation~(\ref{IP-eq-B}). 
This is a subject of the next section.

\setcounter{equation}{0}

\section{Master equation for $\widetilde{\varrho}_S(t)$:
	weak coupling approximation}\label{sec3}

Let us start this section with two assumptions:
\bi
\item
If the subsystem $S$ is small  compared to the environment $B$,
it is reasonable to suppose that, for sufficiently small
time $t$, the state of the environment is close to
$\varrho^{}_{B}(0)$.
Based on the above assumption, we write the solution of
equation~(\ref{IP-eq-B}) as
\beq
\label{IP-rho-B-lin}
\widetilde{\varrho}_{B}(t)\approx \varrho^{}_{B}(0)
- \int^{t}_{0}\rd t'\,\ri{\mathcal L}^{}_{B}(t')\varrho^{}_{B}(0).
\eeq
The above assumption means that we restrict the equation for $\widetilde{\varrho}_B(t)$ by the first order in the interaction.
\item
On the right-hand side of equation~(\ref{IP-eq-S}), we set ${\mathcal U}(t,t')=1$ and
$\widetilde{\varrho}^{}_{B}(t')=\varrho^{}_{B}(0)$,
ensuring the integral terms to be precisely of the 2nd order in interaction.
\ei
Based on these approximations, we get
\beq
\label{IP-eq-S-weak}
\frac{\partial \widetilde{\varrho}^{}_{S}(t)}{\partial t}=
-\ri{\mathcal L}^{}_{S}(t)\widetilde{\varrho}^{}_{S}(t)
-\int^{t}_{0}\rd t'\,
\text{Tr}^{}_{B}
\left\{
{\mathcal L}(t){\mathcal Q}^{(0)}(t'){\mathcal L}(t')
\widetilde{\varrho}^{}_{S}(t')\varrho^{}_{B}(0)
\right\},
\eeq
where
\beq
\label{Q-lin}
{\mathcal Q}^{(0)}(t)A= A
-  \widetilde{\varrho}^{}_{S}(t)\,\text{Tr}^{}_{S}A
- \varrho^{}_{B}(0)\,\text{Tr}^{}_{B}A.
\eeq
After some manipulations with the r.h.s. of equation~(\ref{IP-eq-S-weak}), which are described in detail in the Appendix~A of our recent paper \cite{preprint}, we obtain the final master equation in the interaction picture: 
\bea
\label{master-fin}
& &
\hspace*{-40pt}
\frac{\partial \widetilde{\varrho}^{}_{S}(t)}{\partial t}=
-\ri {\mathcal L}^{(0)}_{S}(t)\widetilde{\varrho}^{}_{S}(t)
-
\int^{t}_{0}\rd t'\,
\left[ R^{}_{S}(t,t'),
\widetilde{\varrho}_{S}(t)-\widetilde{\varrho}_{S}(t')\right]
\nonumber\\
& &
- \int^{t}_{0}\rd t'\,
\left\{
\text{Tr}^{}_{B}
\left\{
{\mathcal L}(t)
{\mathcal L}(t')\widetilde{\varrho}^{}_{S}(t')\varrho^{}_{B}(0)
\right\}
- {\mathcal L}^{(0)}_{S}(t){\mathcal L}^{(0)}_{S}(t')
\widetilde{\varrho}_{S}(t')
\right\},
\eea
where we have used the denotation
\beq
\label{R}
R^{}_{S}(t,t')=\text{Tr}^{}_{B}\left\{
\widetilde{V}(t){\mathcal L}^{}_{B}(t')\varrho^{}_{B}(0)
\right\}=\text{Tr}^{}_{B}\left\{
\left[ \widetilde{V}(t),
\widetilde{V}^{}_{B}(t')\right]\varrho^{}_{B}(0)
\right\}
\eeq
for the kernel of the first integral in equation~(\ref{master-fin}), while the operator ${\mathcal L}^{(0)}_{S}(t)$ is defined as
\beq
\label{L-0-S}
{\mathcal L}^{(0)}_{S}(t)A=\left[\widetilde{V}^{(0)}_{S}(t),A\right],
\qquad
\widetilde{V}^{(0)}_{S}= \text{Tr}^{}_{B}
\left\{
\widetilde{V}(t)\varrho^{}_{B}(0)
\right\}.
\eeq

Some comments on the structure of master equation (\ref{master-fin}) are quite pertinent at this stage:
\bi
\item
The first term on the r.h.s. vanishes if $\varrho^{}_{B}(0)$ is
the \textit{equilibrium} density matrix of the bath and
$$
\text{Tr}^{}_{B}\left\{V\varrho^{}_{B}(0)\right\}=0.
$$
\item
In the above case, the term ${\mathcal L}^{(0)}_{S}(t){\mathcal L}^{(0)}_{S}(t')
\widetilde{\varrho}_{S}(t')$ also vanishes. It is shown in section~\ref{sec4} that the terms with ${\mathcal L}^{(0)}_S$ describe an influence of the initial correlations in the open quantum systems. In particular, the non-integral term contributes to the non-dissipative properties, whereas the integral term represents a new channel of dissipation related to the initial preparation of the system.
\item
The second term in the r.h.s. has some interesting properties:

1) It vanishes in the Markovian limit\footnote{One should distinguish two kinds of approximation for the integrand. The first one looks as
	\linebreak
	$
	\int_0^t \mathcal{K}(t-\tau)f(\tau)\rd\tau\approx
	\int_0^{\mathbf t} \mathcal{K}(t-\tau)\rd\tau f(t).
	$
	This approximation is often called \textit{the MA of the 1st kind} \cite{review} or 
	the \textit{Redfield approximation} (dealt with the corresponding Redfield equation \cite{BP-Book} for the reduced density matrix of the subsystem $S$). 
	The other approximation,
	$
	\int_0^t \mathcal{K}(t-\tau)f(\tau)\rd\tau \approx
	\int_0^{\mathbf{\infty}}\rd\tau \mathcal{K}(t-\tau)\rd\tau f(t)
	$
	is known as \textit{the MA of the 2nd kind} \cite{review}, very often \cite{ZMR} referred to  just as \textit{the Markovian approximation}. In this paper, we use the MA of the 1st kind only.
}, when the dynamics of the kinetic kernel $R_S(t,t')$ is considered fast in comparison with the time evolution of the density matrix; then we set $\widetilde{\varrho}_{S}(t')\approx\widetilde{\varrho}_{S}(t)$.

2) This term is \textit{nonlinear\/} in
$\widetilde{\varrho}_{S}$ since the operator $\widetilde{V}^{}_{B}(t')$
appearing in equation~(\ref{R}) depends on
$\widetilde{\varrho}_{S}(t')$ [see also equation~(\ref{H-int-B})].
A nonlinearity appears due to consideration of the intrinsic dynamics of the environment by means of the time evolution of the reduced density matrix $\widetilde{\varrho}_B(t)$. There is some resemblance with the basic points following from the quantum kinetic theory~\cite{ZMR,MorRop01}, where the running correlations yield nonlinear terms in the kinetic equations, vanishing in the Markovian limit.
\ei

It was mentioned in the end of the previous section that the system of equations (\ref{IP-eq-S})--(\ref{IP-eq-B}) for the reduced density matrices is unmanageable, and obtaining of the master equation for $\widetilde{\varrho}_{S}(t)$ in higher orders in interaction is possible only via some iterative scheme. The eventual master equation would be too cumbersome to be manipulated with. Nevertheless, some possible conclusions can be anticipated by remembering the results which follow from the quantum kinetic theory. For instance, a contribution of the dynamical correlations is known to remain in the Markovian limit \cite{ZMR} if high order approximation in the coupling strength is performed. The origin of the above dynamical correlations \cite{OSID} can be mapped to our concept of the intrinsic bath dynamics. Thus, we could expect the presence of the BD term even in the Markovian limit, as opposed to the above mentioned zeroth contribution following in MA from equation~(\ref{master-fin}).
	
	On the other hand, as it is shown in section~\ref{sec4}, the obtained non-Markovian master equation (\ref{master-fin}), though it is formally of the second order in ${\cal L}(t)$, contains explicitly all the orders in the interaction, if one is going to pass to its convolutionless counterpart. Thus, the latter approach seems to be much easier but not less effective, and generally there is no special need to construct the non-Markovian master equations of higher order in the coupling strength.

Now, it would be instructive to compare the obtained results with those that follow from the traditional ZN projection scheme. In this scheme, one usually uses a different decomposition for the total density matrix instead of (\ref{rho-decomp1}), 
\beq
\widetilde{\varrho}(t)=
\widetilde{\varrho}^{}_{S}(t) \varrho^{}_{B}(0)
+ \delta  \widetilde{\varrho}(t).
\label{Zw-IP-decomp}
\eeq
It is important to note that [c.f. equations~(\ref{IP-Del-rho-prop})]
\beq
\label{Zw-IP-Del-rho}
\text{Tr}^{}_{B}\,\delta  \widetilde{\varrho}(t)=0,
\qquad
\text{Tr}^{}_{S}\,\delta  \widetilde{\varrho}(t)\not=0.
\eeq
In fact, decomposition (\ref{Zw-IP-decomp}) means that one neglects the intrinsic dynamics of the environment by assuming the reduced density matrix of the bath to be equal to its initial value at any time, $\varrho_B(t)\equiv\varrho_B(0)$\footnote{It should be noted that in the original ZN scheme \cite{BP-Book} it is supposed that
	$[H^{}_{B},\varrho^{}_{B}(0)]=0$, i.e., the bath is initially in
	thermal equilibrium. If this initial state differs from thermal
	equilibrium, it is natural to write the decomposition
	of $\varrho(t)$ in the form
		$
	\varrho(t)= \varrho^{}_{S}(t)\varrho^{(0)}_{B}(t)+ \delta \varrho(t)
	$,
	where
		$
	\varrho^{(0)}_{B}(t)= U^{}_{B}(t)\varrho^{}_{B}(0)U^{\dagger}_{B}(t)
	\equiv
	{\rm e}^{-\ri H^{}_{B}t}\varrho^{}_{B}(0){\rm e}^{\ri H^{}_{B}t}.
	$}.
Note that the initial state $\varrho^{}_{B}(0)$ of the environment should not necessarily  be equilibrium: it may contain information about the initial system--bath correlations arising due to the
\textit{selective} \cite{PRA2012} quantum measurements,  as it occurs in the case under study (see section~\ref{sec4}) or \textit{non-selective} measurements~\cite{BP-Book,PRA2015,myPRA2015}.

It is pointed out in \cite{TMF2018} that the ZN projection technique can be viewed as a particular version of the non-equilibrium statistical operator method (NSO), when the interaction term is not taken into account in the relevant distribution $\widetilde{\varrho}_{\rm{rel}}(t)$. The relevant statistical operator is usually considered as the distribution, relating the initial state of the system to the NSO $\widetilde{\varrho}(t)$ through the decomposition
	\bea\label{deltaNSO}
	\widetilde{\varrho}(t)=\widetilde{\varrho}_{\rm{rel}}(t)+\Delta\widetilde{\varrho}(t)
	\eea
	[c.f. equations~(\ref{rho-decomp1}) and (\ref{Zw-IP-decomp})]. Since in the limit $t\to\infty$ the system attains an equilibrium state \cite{ZMR},
	the	correlational part $\Delta\widetilde{\varrho}(t)$ of the total density matrix approaches zero only if the interaction $V$ is included into the relevant distribution. Otherwise, like in the case (\ref{Zw-IP-decomp}) of the ZN projection method and its generalization (\ref{rho-decomp1}), $\lim\limits_{t\to\infty}\delta \widetilde{\varrho}(t)\not=0$ and $\lim\limits_{t\to\infty}\Delta \widetilde{\varrho}(t)\not=0$,
	since the equilibrium distribution of the composite system is not factorisable in the operators, belonging to the different subsystems $S$ and $B$. Here, we leave aside all the peculiarities dealing with  the system tending to equilibrium, and refer our readers to the papers~\cite{TMF2018, OSID}. 
	
	Secondly, the expression (\ref{Zw-IP-Del-rho}) renders the main disadvantage of the ZN projection method quite obvious: the state of the environment is postulated 
	to be described by an equilibrium distribution plus some correction $\mbox{Tr}_S\delta\widetilde{\varrho}(t)$, which seems reasonable only for the case of small interactions. On the contrary, the generalization of the ZN method implies that the reduced density matrix of the environment has its own dynamics, completely incorporated in the distribution  $\widetilde{\varrho}_B(t)$, see equations~(2.13) and (2.17) in \cite{CMP}, whereas approximation (\ref{IP-rho-B-lin}) can be treated as its linearized version.

In the weak coupling limit (see Appendix B in \cite{preprint} for details), 
the ZN master equation looks as follows:
\bea
\label{ZW-master-wI}
\frac{\partial \widetilde{\varrho}^{}_{S}(t)}{\partial t}=
-\ri{\mathcal L}^{(0)}_{S}(t)\widetilde{\varrho}^{}_{S}(t)
{}- \int^{t}_{0}\rd t'\,
\left\{
\text{Tr}^{}_{B}
\left\{
{\mathcal L}(t)
{\mathcal L}(t')\widetilde{\varrho}^{}_{S}(t')\varrho^{}_{B}(0)
\right\}
- {\mathcal L}^{(0)}_{S}(t){\mathcal L}^{(0)}_{S}(t')
\widetilde{\varrho}_{S}(t')
\right\}.
\eea
	Note that the above reasoning should not be misleading about a formal resemblance of the equation for the density matrix of the environment $\widetilde{\varrho}_B(t)=
	\varrho^{}_{B}(0)
	+\mbox{Tr}_S \delta\widetilde{\varrho}(t)$,
	which follows from the relations (\ref{Zw-IP-decomp})--(\ref{Zw-IP-Del-rho}) of the ZN method, and the approximation (\ref{IP-rho-B-lin}) in our generalized projecting scheme. The final results for the master equations (\ref{master-fin}) and (\ref{ZW-master-wI}) are completely different, since they are derived starting from two different basic points: there is only one equation for the reduced density matrix of the open quantum system in the framework of the ZN projection technique, while taking the BD into consideration one deals with the coupled chain of equations for $\widetilde{\varrho}_S(t)$ and $\widetilde{\varrho}_B(t)$, see equations~(\ref{IP-eq-S})--(\ref{IP-eq-B}). As a consequence, in contrast to equation~(\ref{master-fin}), the master equation (\ref{ZW-master-wI})
	does not contain a nonlinear term with $R^{}_{S}(t,t')$ due to  the intrinsic dynamics of the environment being excluded from consideration. Nevertheless, in the Markovian limit,
	both master equations are identical. This is quite obvious, since the effect of BD [considered in terms of $\widetilde{\varrho}_B(t)$] is expected to manifest itself only in the initial stage of the system evolution. 

To check out the consistency and robustness of the generalized projection scheme, one should use the master equation (\ref{master-fin}) to derive the quantum kinetic equations for the observables in some simple models (preferably, exactly solvable ones). This is done in the next section, where we obtain a dynamic equation for coherence, considering the purely dephasing model \cite{PRA2012}.

\section{Non-Markovian quantum kinetic equation for	a dephasing model}\label{sec4}

\setcounter{equation}{0}

We consider a simple version of a spin-boson model describing the
two-state system ($S$) coupled to the bath ($B$) of harmonic
oscillators \cite{PRA2012,TMF}. In the spin representation for a
qubit, the total Hamiltonian of the model is written as
\bea\label{H}H=H_S+H_B+V=\frac{\omega_0}{2}\sigma_3+\sum\limits_k\omega_k
b^{\dagger}_k b_k+\sigma_3\sum\limits_k(g_k b^{\dagger}_k+g^*_k
b_k),\eea where $\omega_0$ is the energy difference between the
excited state $|1\rangle$ and the ground state $|0\rangle$ of the
qubit, and $\sigma_3$ is the 3rd Pauli matrix $\sigma_i$, $i=\{1, 2 , 3\}$. Bosonic creation and annihilation
operators $b^{\dagger}_k$ and $b_k$ correspond to the $k$-th bath
mode with the frequency $\omega_k$.

Since $\sigma_3$ commutes with Hamiltonian (\ref{H}), it does not
evolve in time, $\sigma_3(t)=\sigma_3$. Hence, the interaction
operator can be easily expressed as
\bea\label{tildeV1}\widetilde V(t)\equiv\sigma_3\sum\limits_k
{\cal F}_k (t)=\sigma_3\sum\limits_k\left\{g_k(t)
b^{\dagger}_k+g^*_k(t)b_k \right\},
\qquad
g_k(t)=g_k \exp(\ri \omega_k t).
\eea

Let us introduce in a usual fashion the spin
inversion operators $\sigma_{\pm}=(\sigma_1\pm i\sigma_2)/2$,
which obey the permutation relations
$\left[\sigma_3,\sigma_{\pm}\right]=\pm 2\sigma_3$. Our task is to
obtain the quantum kinetic equations for the mean values
$\langle\widetilde{\sigma}_{\pm}\rangle_{S}^{t}=\mbox{Tr}_S\{\widetilde\rho_S(t)\sigma_{\pm}\}$ dealing with the
system coherence. For simplicity, we consider the case with
$\langle\widetilde{\sigma}_+\rangle_{S}^{t}$, since the equation for $\langle\widetilde{\sigma}_-\rangle_{S}^{t}$ can be easily obtained in a similar manner.

Using the basic equation (\ref{master-fin}) for the reduced density
matrix $\widetilde\rho_S(t)$ and taking a trace over the system
variables, one can derive the following kinetic equation:
\bea\label{kinet1}\nonumber \frac{\partial
	\langle\widetilde\sigma_{+}\rangle_{S}^{t}}{\partial
	t}=
\ri\left\langle\left[\widetilde V_S^{(0)}(t),\widetilde\sigma_+\right]\right\rangle^t_{S}
+\int\limits_0^t
\rd t'\biggl\{
\left\langle\left[R_S(t,t'),\widetilde\sigma_+\right]\right\rangle^t_{S}-
\left\langle\left[R_S(t,t'),\widetilde\sigma_+\right]\right\rangle^{t'}_{S}
\biggr\}\\
-\int\limits_0^t \rd t'\biggl\{\left\langle\left\langle\left[\widetilde
V(t),\left[\widetilde
V(t'),\widetilde\sigma_+\right]\right]\right\rangle_B\right\rangle^{t'}_{S}+
\left\langle\left[\widetilde
V_S^{(0)}(t),\left[\widetilde
V_S^{(0)}(t'),\widetilde\sigma_+\right]\right]\right\rangle^{t'}_{S}
\biggr\},\eea
where we denote the bath averaging as $\langle\ldots\rangle_{B}\equiv\mbox{Tr}_B\left\{\varrho_{B}(0)\ldots
\right\}$.

Having calculated all commutators in equation~(\ref{kinet1}) and having performed thermal averaging (see Appendixes C and D in \cite{preprint} for details), we can write down the final kinetic equation for the generalized coherence:
\bea\label{kinetFin}
\frac{\rd\left\langle\widetilde\sigma_+\right\rangle^t_{S}}{\rd t}&=&\ri
A_{\rm{init}}\int\limits_0^{\infty}\frac{J(\omega)}{\omega}\cos\omega
t\,\rd\omega\left\langle\widetilde\sigma_+\right\rangle^t_{S}\nonumber\\
&-&
\ri\langle\sigma_3\rangle\int\limits_0^t\Biggl\{\int\limits_0^{\infty}J(\omega)\sin[\omega(t-t')]\rd\omega\Biggr\}\left(\langle\widetilde\sigma_+\rangle^{t}_{\widetilde
	S}-\langle\widetilde \sigma_+\rangle^{t'}_{S}\right)\rd t'
\nonumber\\
&-&\int\limits_0^t\Biggl\{\frac{1}{2}\int\limits_0^{\infty}J(\omega)\coth\left(\frac{\omega}{2
	k_{\text B}
	T}\right)\cos[\omega(t-t')]\rd\omega
\nonumber\\
&-&2(A_{\rm{init}}^2-1)\int\limits_0^{\infty}\frac{J(\omega)}{\omega}\cos\omega
t\,\rd\omega\int\limits_0^{\infty}\frac{J(\omega')}{\omega'}\cos\omega't'\,\rd\omega'\Biggr\}\left\langle\widetilde\sigma_+
\right\rangle^{t'}_{S}\rd t',\eea
where the quantity
\bea\label{A}  A_{\rm{init}}=
\frac{\exp(\beta\omega_0/2)(1-\langle\sigma_3\rangle)-\exp(-\beta\omega_0/2)(1+\langle\sigma_3\rangle)}{\exp(\beta\omega_0/2)(1-\langle\sigma_3\rangle)+
	\exp(-\beta\omega_0/2)(1+\langle\sigma_3\rangle)}\eea
is related to the initial correlations (in other words, it deals with the initial preparation of the system due to some kind of the quantum measurement \cite{PRA2012,PRA2015}).

In the kinetic equation (\ref{kinetFin}), the qubit-bath coupling is modelled by the spectral weight function $J(\omega)$ taken in its standard form \cite{PRA2012}
\bea\label{J1(w)}
J(\omega)=\lambda\Omega^{1-s}\omega^s \exp(-\omega/\Omega).
\eea

The parameter $\lambda\sim |g_k|^2$ in equation~(\ref{J1(w)}) denotes a dimensionless coupling constant, while $\Omega$ is the cut-off frequency. Depending on the exponent $s$, we can distinguish three coupling regimes: the sub-Ohmic at $0<s<1$, the Ohmic at $s=1$, and the super-Ohmic at $s>1$. 

Let us discuss some peculiarities of the quantum
kinetic equation (\ref{kinetFin}). First of all, we recall that
the second term in the r.h.s. of (\ref{kinetFin}) vanishes in the Markovian
limit. On the other hand, we can always pass from the mean values in
the interaction picture to the averages taken with the statistical
operator $\rho_S(t)$ according to the rule
$\left\langle\widetilde\sigma^+\right\rangle^t_{
	S}\rightarrow\exp(-\ri\omega_0
t)\left\langle\sigma^+\right\rangle^t_S$. Then, the second term in the r.h.s. of (\ref{kinetFin}), being an imaginary one, along
with the first (non-dissipative or ``quasi-free'') term should add up to $\omega_0t$, yielding the corresponding phase shift
\cite{PRA2012}.

Secondly, the initial state of the qubit contributes to the quasi-free term as well as to the last summand
in the r.h.s. of equation~(\ref{kinetFin}) via $A_{\rm{init}}$.
Besides, while the kinetic kernels in the 2nd and the
3rd terms have a usual form ${\cal
	K}(t-t')$, this is not the case for the last term dealing with the
initial correlations in the ``qubit+bath'' system. This is not surprising: the dynamics of initial correlations is not a stationary process with a typical convolution dependence; the initial correlations decay in time due to the aging effect \cite{Pottier}.
Thirdly, the contribution of the initial correlations is of the 4th order in interaction.

In our recent paper \cite{preprint} we  solved the generic non-Markovian equation (\ref{kinetFin}) numerically. We  considered the Ohmic and super-Ohmic (with $s=3/2$) qubit-bath coupling and studied the qubit dynamics at both low and high temperatures. The results  were not quite satisfactory from the physical point of view:  the BD being taken into account led to an uncontrollable increase of $\mbox{Im}\langle\widetilde{\sigma}_+(t)\rangle$ at low temperatures. The picture was not improved much even at high temperatures, since the imaginary part of the system coherence tended to saturation. This contradicts the conclusions made in \cite{PRA2012} that the partial decoherence is admitted only at the strong super-Ohmic regime with $s>2$, while at smaller values of the ohmicity index the system coherence always tends to zero at large times. Since no systematic analysis can be proposed in the non-Markovian regime but only the possibility to solve the kinetic equations numerically, and since the reasons leading to the above mentioned unphysical behaviour of the coherence were formulated only qualitatively, we perform the MA\footnote{Formally, it corresponds to the ZN approach in a time convolutionless picture.} of equation~(\ref{kinetFin}) to obtain exact analytical expressions and to study the system dynamics more carefully.

The first term in the the r.h.s. of equation~(\ref{kinetFin}) is proportional to the time derivative of the phase shift,
\bea\label{chi}
\chi(t)=\frac{\sinh(\beta\omega_0/2)-\langle\sigma_3\rangle\cosh (\beta\omega_0/2)}{\cosh(\beta\omega_0/2)-\langle\sigma_3\rangle\sinh (\beta\omega_0/2)}\Phi(t),\qquad
\Phi(t)=\int\limits_0^{\infty}\rd\omega J(\omega)\frac{\sin\omega t}{\omega^2},
\eea
which is nothing else but the linearized over $\Phi(t)$ version of the expression (32) from \cite{PRA2012}. Furthermore, taking the generalized coherence out of the integral in the third term and integrating the kinetic kernel over $t'$, we come to the conclusion that it deals with the time derivative of 
\bea
&&\gamma(t)=\gamma_{\rm{vac}}(t)+\gamma_{\rm{th}}(t), \nonumber\\
&&\gamma_{\rm{vac}}(t)=\int\limits_0^{\infty}d\omega J(\omega)\frac{1-\cos\omega t}{\omega^2},\qquad
\gamma_{\rm{th}}(t)=\int\limits_0^{\infty}d\omega J(\omega)\left[\coth(\beta\omega/2)-1
\right]\frac{1-\cos\omega t}{\omega^2}, \label{gamma_vac_th}
\eea
where $\gamma_{\rm{vac}}(t)$ ($\gamma_{\rm{th}}(t)$) denotes the vacuum (thermal) contribution to the generalized decoherence. 

In a similar way, performing the MA, we verify that the last term of equation~(\ref{kinetFin}) is related to the time derivative of the correlational contribution to the generalized decoherence,
\bea\label{gamma_cor}
\gamma_{\rm{cor}}(t)=\frac{1-\langle\sigma_3\rangle^2}{2[\cosh(\beta\omega_0/2)-\langle\sigma_3\rangle\sinh (\beta\omega_0/2)]^2}\, \Phi(t)^2.
\eea
Again, the expression (\ref{gamma_cor}) is nothing else but the series expansion of equation~(31) from \cite{PRA2012} up to the fourth order in interaction.

Now, let us examine more thoroughly the role of the intrinsic BD, which is described by the second term in the r.h.s. of the kinetic equation (\ref{kinetFin}). 
This term has a very specific structure: it does not depend on the temperature {\bf in spite of having been generated by the qubit environment.} In some sense, it is similar to the vacuum contribution $\gamma_{\rm{vac}}(t)$ to the decoherence function. It is obvious that the above similarity is rather coincidental, being a cumulative effect of the applied approximation (\ref{IP-rho-B-lin}) for $\tilde\rho_B(t)$ and a specific feature (non-ergodicity) of the dephasing model. To get some more information about the role of BD, let us
go beyond the MA and expand the difference of the generalized coherences up to the first order in $\tau=t-t'$: 
\bea\label{expandSig+}\langle\widetilde\sigma_+\rangle^{t}_{
	S}-\langle\widetilde \sigma_+\rangle^{t'}_{S}=\frac{\rd\langle\widetilde\sigma_+\rangle^{t}_{
		S}}{\rd t} (t-t')+o(t-t')^2.\eea 
Since the time derivatives of both the phase shift and the generalized coherence are of the 2nd order in the coupling constant, the second term in the r.h.s. of (\ref{kinetFin}) is proportional to $|g_k|^4$.
Thus, taking (\ref{chi})--(\ref{expandSig+}) into account,  it is straightforward to solve equation~(\ref{kinetFin}) and to obtain the generalized coherence in its usual form
\bea\label{solSigmaP}
\langle
\widetilde{\sigma}_+
\rangle^t_S=\langle\widetilde{\sigma}_+
\rangle^{t=0}_S \exp[\ri\bar{\chi}(t)-\bar\gamma(t)].
\eea
Here, the renormalized values of the phase shift $\bar\chi(t)$ and the generalized decoherence $\bar\gamma(t)$ up to the 4th order in interaction read as follows:
\bea\label{renormChi}
\bar\chi(t)=\chi(t)+\int\limits_0^t\, \rd t' F(t')\frac{\rd[\gamma_{\rm{vac}}(t')+\gamma_{\rm{th}}(t')]}{\rd t'},
\eea	
\vspace*{-5mm}
\bea\label{renormGamma}
\bar\gamma(t)=\gamma_{\rm{vac}}(t)+\gamma_{\rm{th}}(t)+\gamma_{\rm{cor}}(t)-\int\limits_0^t\, \rd t' F(t')\frac{\rd \chi(t')}{\rd t'},
\eea
where the function
\bea\label{intSin}
F(t)=\langle\sigma_3\rangle\int\limits_0^t \rd\tau\, \tau\left\{\int\limits_0^{\infty}J(\omega)\sin\omega\tau\,\rd\omega
\right\}
\eea
can be calculated explicitly, but is too cumbersome to be presented here. In the integrand of (\ref{renormChi}) we  omitted  $\gamma_{\rm{cor}}(t)$, since it is already of the 4th order in the coupling constant. At the chosen order in the coupling constant, the BD contribution to the decoherence function vanishes for the initially uncorrelated subsystems [see the equation~(\ref{renormGamma})], since the phase shift is zero in such a case \cite{PRA2012}. It can be said that the initial correlations and the intrinsic BD are working together in the composite quantum system to enhance or decrease the coherence, depending on the sign of $\langle\sigma_3\rangle$. 
Since in the dephasing model the exact value of the correlational contribution to the decoherence function is known \cite{PRA2012},
	\bea\label{gamCorExact}
	\gamma_{\rm{cor}}^{\rm{exact}}=-\frac{1}{2}\ln\left\{1-\frac{(1-\langle\sigma_3\rangle^2)\sin^2[\Phi(t)]}{[\cosh(\beta\omega_0/2)-\langle\sigma_3\rangle\sinh(\beta\omega_0/2)]^2}\right\},
	\eea
	it is reasonable to treat the last term in (\ref{renormGamma}) as a renormalization of (\ref{gamma_cor}) and to compare
	\vspace{-3mm}
	\bea\label{renormGammaCor}
	\bar\gamma_{\rm{cor}}(t)=\gamma_{\rm{cor}}(t)-\int\limits_0^t\, \rd t' F(t')\frac{\rd\chi(t')}{\rd t'}
	\vspace{-7mm}
	\eea
	with the above exact result (\ref{gamCorExact}).
	
\begin{figure}[!t]
		\centerline{\includegraphics[height=0.25\textheight,angle=0]{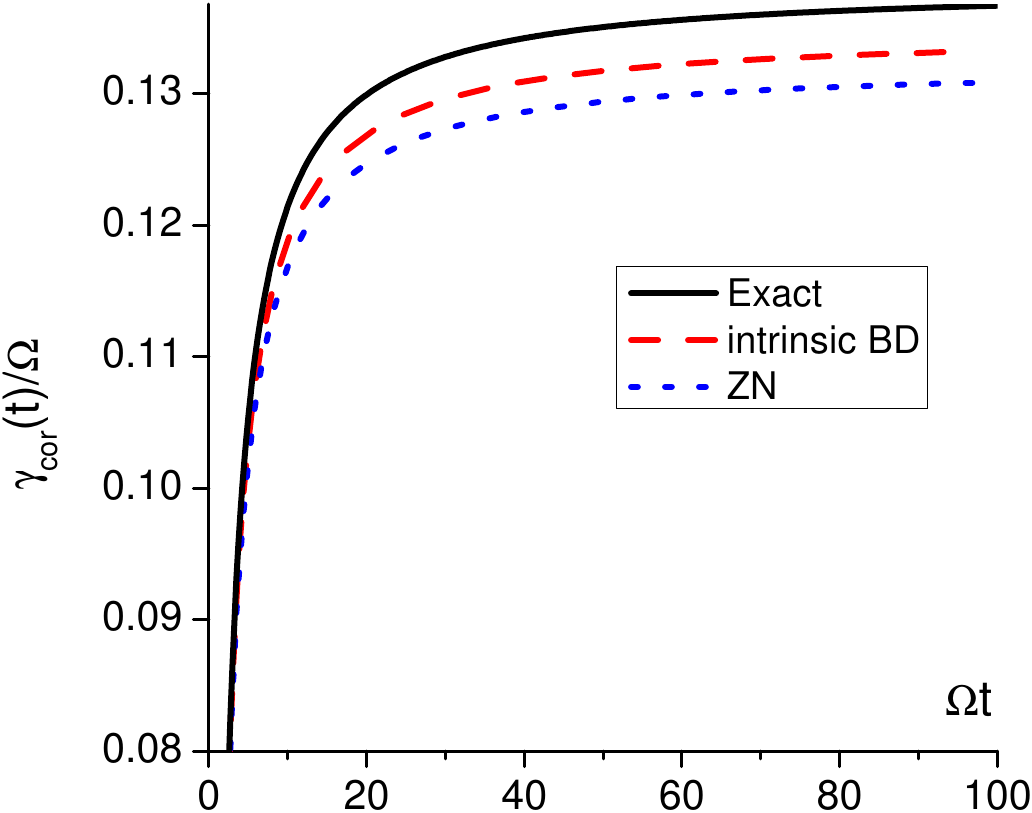}}
		\caption{Time dependence of the correlational contribution to the generalized decoherence, Ohmic coupling. Model parameters: $\omega_0/\Omega=1$, $\beta\omega_0=0.1$, $\langle\sigma_3\rangle=0.2$, $\lambda=1/3$. The solid, dashed and dotted curves 
				correspond to the exact (\ref{gamCorExact}), renormalized (\ref{renormGammaCor}) and the ZN (\ref{gamma_cor}) expressions.}
			\label{Fig1}
	\end{figure}

\begin{figure}[!t]
		\centerline{\includegraphics[height=0.25\textheight,angle=0]{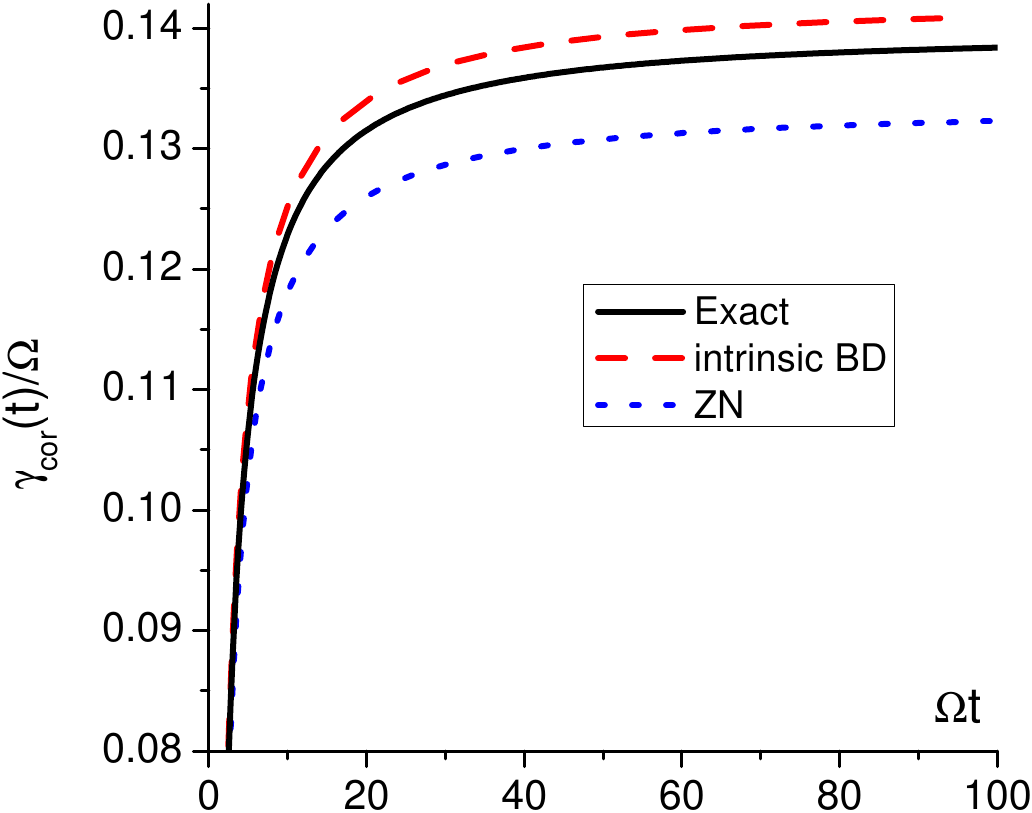}}
		\caption{The same as in figure~\ref{Fig1} at $\beta\omega_0=5$ and  $\langle\sigma_3\rangle=0.99$.}
		\label{Fig2}
	\end{figure}

	As it is seen in figures~\ref{Fig1} and \ref{Fig2}, there is a domain of the state parameters (the temperature $1/\beta$ and the inversion of the mean levels populations  $\langle\sigma_3\rangle$), where the renormalized values (\ref{renormGammaCor}), obtained within our generalized projection technique, reproduce the exact result (\ref{gamCorExact}) better than those derived using the ZN approach~(\ref{gamma_cor}). This region is found to be characterized by the values $-0.2<A_{\rm{init}}<0$, meaning a slight qubit frequency downshift. 
	\noindent
	It should be noted  that such a behaviour of $\gamma_{\rm{cor}}(t)$ is typical both in the high (figure~\ref{Fig1}) and low (figure~\ref{Fig2}) temperature limits. 
	It could be also shown that in the moderate super-Ohmic regime, the obtained dynamics shows tendencies similar to those presented at $s=1$. However, at the sub-Ohmic coupling, the behaviour of $\gamma_{\rm{cor}}^{\rm{exact}}(t)$ is known to be non-monotonous in time \cite{PRA2012}, and none of the approximations (\ref{gamma_cor}) or (\ref{renormGammaCor}) can reproduce the true dynamics of the correlational contribution to the decoherence function at large times due to a specific kind of coupling \cite{review}.

	On the other hand, the lowest series expansion in (\ref{expandSig+}) does not allow us to  reliably estimate the correction to the phase shift according to equation~(\ref{renormChi}), it arises even at $\chi(t)=0$, contradicting the well-known statement \cite{PRA2012}, which attributes the qubit frequency renormalization solely to the appearance of the initial correlations in the composite $S+B$ system. Going beyond the 4th order in the interaction in calculations of the renormalized values of $\chi(t)$ and $\gamma_{\rm{cor}}(t)$ could eventually improve the situation. However, it also implies taking into account the next terms in the series expansion (\ref{expandSig+}) which automatically would deviate the expected result from the exact exponential form (\ref{solSigmaP}).
	
	Nevertheless, the results presented in figures~\ref{Fig1} and \ref{Fig2} show that the proposed generalization of the projection technique can be considered as a good alternative to other perturbative methods of solving the quantum master equations \cite{BP-Book}, which perceptibly improves the ZN results already in the lowest order in the coupling strength.

	To conclude this section, we mention that all the results are obtained for the system that was initially prepared by the \textit{selective} quantum measurement. The quantity (\ref{A}), entering all the expressions dealing with initial correlations, is uniquely defined by the above preparation measurement. However, there is a much more interesting situation, when the system is prepared due to a \textit{non-selective} measurement scheme~\cite{BP-Book}. In such a case, not only the decoherence induced by the initial correlations is observed but also the inverse process (recoherence, or the coherence enhancement) is possible. In \cite{PRA2015}, we proposed some kinds of measurements leading to the permanent coherence growth at small times $t\sim 1/\Omega$. Obviously, at times $t\gg 1/\Omega$, the other dissipation channels (thermal and vacuum ones) prevail, yielding a decrease of the system coherence. There is also a possibility to prepare the system in such an initial stage, when the periods of recoherence and decoherence  alternate. In the very recent papers \cite{Giraldi2020,Giraldi2021}, the evolution of quantum coherence for the same dephasing model is studied in detail over short and long times.
	
	Thus, we are in a position to perform a similar analysis of the BD contribution to the system, which is initially prepared after the \text{non-selective} quantum measurements, since the general ideas leading to the ``renormalized'' functions (\ref{renormChi})--(\ref{renormGamma}) are valid in this case as well, and the general expression (\ref{solSigmaP}) for the system coherence remains unaltered. However, the expression for the ``renormalized'' correlational contribution $\bar\gamma_{{\rm cor}}(t)$ to the decoherence/recoherence rate would consist of too many parameters to be analyzed reliably. As far as our case of \textit{selective} preparation measurements is concerned, the short or long time analysis of the renormalized function (\ref{renormGammaCor}) can hardly be informative or useful, because of its small contribution to the (permanent!) system decoherence. Therefore, we prefer to focus our efforts on the investigation of a measure of its deviation from the exact result (\ref{gamCorExact}) in order to verify the validity and robustness of our generalized projection scheme beyond the Zwanzig-Nakajima framework.

\section{Conclusions and outlook}\label{sec5}

In this paper, we generalize the ordinary ZN projection scheme in order to derive a master equation for the open quantum system weakly coupled with its surrounding. We start from the chain of equations for the density matrices of the $S$- and $B$-subsystems and obtain a non-Markovian master equation for $\widetilde{\varrho}_S(t)$. 
We simplify this equation, restricting ourselves by the lowest approximation in the coupling constant. 
The consideration of the intrinsic BD yields an extra term in the master equation, which is found to be nonlinear in $\widetilde{\varrho}_S(t)$ and vanish in the Markovian limit.  When  the dynamic equation for $\widetilde{\varrho}_B(t)$ is neglected, we come to the standard ZN projecting scheme \cite{BP-Book}.

In order to test the developed method we  applied it to the open quantum system described by a very simple dephasing model. The non-Markovian quantum kinetic equation for the ge\-ne\-ra\-li\-zed coherence has been derived. In the MA, which means nothing else but the time convolutionless ZN result, its solution up to the 2nd order in interaction coincides with the exact one obtained in \cite{PRA2012}.
	We  also analyzed the kinetic kernel related to the BD, which is found to be temperature 
	independent. When going beyond the MA in the above mentioned term, we obtain the renormalized values of the phase shift and the generalized decoherence. We  compared the renormalized correlational contribution to the decoherence function with the exact result \cite{PRA2012} and with that of the ZN approach. It is found that consideration of the intrinsic BD brings the corresponding curves closer to the exact data  compared to the results obtained within the ordinary ZN projection scheme.
	
However, some model peculiarities, e.g., its non-ergodicity do not allow us to take the full advantages of our method:
in the lowest order in interaction, the kernel dealing with the BD, turns out to be temperature independent, which seems a bit strange if one speaks about the effect of the qubit surrounding. Other difficulties arise from the necessity of taking the higher order terms into consideration which inevitably causes significant computational complications\footnote{In this context it should be noted that a formal application of the perturbation theory in the interaction can sometimes be misleading: in the case of the dephasing model under study, any eventual higher-order corrections to the vacuum $\gamma_{\rm{vac}}(t)$ and thermal $\gamma_{\rm{th}}(t)$ terms should be rejected as contradicting the exact solution.}
	and, from the mathematical point of view,  the solution for the generalized coherence can even deviate from its exact exponential form \cite{PRA2012}.

There is also a different approach to treat the influence of BD, which is based on the consideration of dynamical correlations in a composite system \cite{particles}. These long-living dynamical correlations, which are associated with the total energy conservation, play an important role in the transition to the Markovian regime and the subsequent equilibration of the system. In our recent paper \cite{OSID}, we  applied this approach to the dephasing model and obtained nonlinear kinetic equations, involving the so-called correlational temperature. Though the above method does not admit consideration of the intrinsic BD explicitly, it allows one to study the build-up of the dynamic correlations between the qubit and the bath, as well as its time evolution from the initial stage up to the system equilibration. It is interesting that the dynamic correlations act in a similar way, renormalizing the qubit frequency. Thus, we dare to say that the studies of the BD influence on the state of open quantum systems, the build-up of the system--bath correlations, as well as a choice of the most suitable models for verification of these approaches, are still far from completeness and should be continued in the future.

\section*{Note added in proof}
My attention (V.~V.~I.) has been drawn to a very recent paper \cite{PRX2020} where the problems of intrinsic bath dynamics were studied in the framework of the so-called correlation picture, which connects a correlated bipartite state to its uncorrelated counterpart. Some of the results and conclusions are very close to ours, obtained in the current paper and previously in  \cite{preprint}, namely: i) generalization of the projecting operators; ii) the chain of equations for the reduced density matrices of the system and the bath. However, there are also some distinctive points: i) the models utilized in \cite{PRX2020} in order to verify the robustness of their approach differ from ours; ii) we have separated 
the initial correlations and the BD contributions, yielding different forms of the corresponding time-convolutionless master equations.
Similar results, obtained independently by different approaches, undoubtedly confirm the importance of dynamical correlations in the open quantum systems and the necessity to take them into account.


\newpage

\ukrainianpart

\title{Декогерентність у відкритих квантових системах: вплив внутрішньої динаміки термостату}
\author{В. В. Ігнатюк\refaddr{label1}, \framebox{В. Г. Морозов}\refaddr{label2}}
\addresses{
\addr{label1} Інститут фізики конденсованих систем НАН України, вул. Свенціцького, 1, 79011 Львів, Україна
\addr{label2} МІРЕА - Російський технологічний університет, просп. Вернадського, 78,  119454 Москва, Росія
}

\makeukrtitle

\begin{abstract}
\tolerance=3000%
Шляхом узагальнення стандартної проекційної техніки Цванцига-Накаджими (ЦН) отримано немарківське керуюче рівняння для відкритої квантової системи. З цією метою була записана система рівнянь для приведених матриць густини термостату $\varrho_{B}(t)$ та системи $\varrho_{S}(t)$. Формальний розв'язок рівняння для $\varrho_{B}(t)$ у другому порядку за взаємодією приводить до появи досить специфічного додаткового члена, пов'язаного з власною динамікою термостату. Цей доданок є нелінійним стосовно приведеної матриці густини системи $\varrho_{S}(t)$ та зануляється у марківській границі. Для перевірки надійності запропонованого підходу, узагальнення методу ЦН застосовано до простої моделі з розфазуванням. Отримані кінетичні рівняння досліджувались як у марківському наближенні, так і при виході за його межі (стосовно доданку, зв'язаного з внутрішньою динамікою термостату) та порівнювались з точними результатами.
\keywords {керуюче рівняння, проекційна техніка, модель з розфазуванням, декогерентність}

\end{abstract}

\end{document}